\documentclass[11pt, oneside]{article}   	% use "amsart" instead of "article" for AMSLaTeX format
\usepackage{geometry}                		% See geometry.pdf to learn the layout options. There are lots.
\usepackage{graphicx}				% Use pdf, png, jpg, or eps§ with pdflatex; use eps in DVI mode
\usepackage{amsmath}
\usepackage{amssymb}
\usepackage{amsthm}
\newtheorem{theorem}{Theorem}
\newtheorem{proposition}[theorem]{Proposition}
\newtheorem{remark}[theorem]{Remark}
\newcommand{\inv}{\mathrm{inv}}
\newcommand{\sd}{\partial}
\newcommand{\Z}{\mathbb{Z}}

\newtheorem{corollary}[theorem]{Corollary}

\newtheorem{lemma}[theorem]{Lemma}
\newcommand{\C}{\mathbb{C}}

\newcommand{\XA}{X_*(A)}

\newcommand{\Ext}{\mathrm{Ext}}

\newcommand{\Tor}{\mathrm{Tor}}
\newcommand{\Hom}{\mathrm{Hom}}

\newcommand{\Id}{\mathrm{id}}

\newcommand{\cF}{\mathcal{F}}

\title{Quantum deformation of planar amplitudes }
\author {M. Movshev, A . Schwarz}
\date{}
\begin {document}
\author{M. V. Movshev\\Stony Brook University\\Stony Brook, NY 11794-3651, USA
\\ A. Schwarz\\ Department of Mathematics\\ 
University of 
California \\ Davis, CA 95616, USA}
\maketitle

\vskip .3in
It was realized recently that on-shell diagrams are objects of fundamental importance in the analysis of scattering amplitudes. In four-dimensional theories planar on-shell diagrams are closely related to the positive Grassmannian and the cell decomposition of it into the union of so called positroid cells \cite{A}. 
(The most complete results were obtained for $\mathcal{N}=4$ - supersymmetric Yang-Mills theory, but the relation to positive Grassmannian exists also in other cases.) From the other side it was discovered  \cite {LLR} that there exists a bijective correspondence between positroid cells in positive Grassmannian   and   prime ideals in quantum Grassmannian that are invariant with respect to the torus action. (This Grassmannian was constructed in the framework of quantum group theory.) We come to the idea that the theory of \cite {A} can be quantized (more precisely, $q$-deformed); one can hope that the quantization can lead to physically interesting results. This idea is supported by the fact  that the  Grassmannian can be considered as cluster variety \cite{FGA} with compatible Poisson structure and, as was shown in \cite {FG},\cite {FGQD} in this situation there exists a unique quantization procedure that should lead to the same quantum Grassmannian.
							% Activate to display 
We  prove some  results in this direction. Namely, we  establish that volume forms on positroids used in \cite{A} can be $q$-deformed to Hochschild homology classes of corresponding quantum algebras. (It is well known that Hochschild homology classes can be considered as non-commutative deformation of differential forms. Notice, however, that so called twisted Hochschild homology  also can be regarded as deformation of differential forms; our results can be applied to twisted Hochschild homology.)

This fact follows from  much more general results that can be applied to quantum cluster algebras in the sense of \cite {BZ}  and to quantum spaces in the sense of \cite{FGD}. 

We use the deformed volume form to  $q$- deform the contribution of a positroid cell to scattering amplitude. Our construction  can be used to simplify calculations also in the case of classical Grassmannian ( see Section 7).

The scattering amplitude can be described as a sum of contributions of some positroid cells. The nicest description of the amplitude can be given in terms of amplituhedron \cite {AT}.

We hope  that one can "quantize" results of \cite {AT} , \cite {AHT} applying our results to partial flag varieties. (It was shown in \cite{FGA} that partial flag varieties can be considered as cluster varieties, see also \cite {GLS}.) The partial flag varieties arise naturally in the framework of \cite {A} either as flags of subspaces of dimensions $2, m, n-2$ or , in twistor formalism, as flags of subspaces of dimension $4,m.$

The physical interpretation of  our results is not clear. It is possible that they are related  to non-commutative gauge theories in the spirit of \cite {CDS}, \cite {KS}, \cite {SW}. Another conjecture: quantizing planar on-shell diagrams with $q=e^{i\hbar}$ equal to the $N$-th root of unity (in other words $\hbar =\frac{2\pi}{N}$) we can obtain physical quantities associated with    
gauge theories with the gauge group $U(N).$  Here $q$ is the parameter entering the definition of the quantum Grassmannian. (Recall that the planar diagrams correspond to the limit $N\to \infty.$)  Due to existence of Seiberg-Witten map \cite {SW} these two conjectures can be true simultaneously; if this is the case the $U(N)$ gauge theory  should interact with constant $B$-field. ( Interaction with $B$-field  can explain the violation of Lorentz invariance that should be present in $U(N)$ gauge theory.)   

The second conjecture is supported by the remark that a similar statement is true for Chern-Simons theory. It is well known that Chern-Simons theory with gauge group $U(N)$  is equivalent to topological string  with string coupling constant $g_s=\frac{2\pi}{k+N}$ \cite {Wit}, \cite {MAR}. (Here $k$ stands for the coefficient in front of Chern-Simons action functional-the level.)  From the other side the topological string for any genus can be obtained from genus zero topological string by means of quantization and the string coupling constant plays the role of Planck constant \cite {WITT}. This agrees with our conjecture. ( Genus zero string theory corresponds to $N=\infty$ and planar diagrams.)

One more reason to expect that after the quantization of Grassmannian we should get $U(N)$ gauge theory is based on  the following remark.  The scattering amplitudes of $s$ particles in $N=\infty$ case are functions on some manifold. (For example, in twistor formalism one can consider them  as functions on the configuration space of $s$ points on the manifold $\mathbb{C}P^{3|4}$.) After quantization  of this manifold with $q$ obeying $q^N=1$  we should get functions depending of the points of the same manifold and additionally of some discrete variables. (This follows from the fact that for  $q^N=1$ the quantized algebra has a large center isomorphic to the original commutative algebra.)  This agrees with the picture of $U(N)$ gauge theory where these discrete variables can be identified with $N^2$ "colors" for every particle. It is a non-trivial check of our conjecture that the number of discrete variables is precisely $N^{2s}$ where $s$ is the number of particles; under some  assumptions we were able to confirm this statement.

Our considerations are based on consideration of quantum cluster algebras in the sense of \cite {BZ} and of quantum spaces in the sense of \cite {FG}, \cite {FGQD}. However, we do not need the full strength of the theory of cluster algebras and our exposition does not depend on this theory. We introduce  much simpler notions of poor man cluster algebra and poor man quantum space that are sufficient for our goals. 

We will work with unital associative algebras over $\mathbb {C}.$ Prime ideal of   an algebra (or of a ring)  $\cal A$ is a two-sided  ideal $\cal I\neq \cal A$ such that for two ideals $X,Y$  obeying $XY\subset \cal I$ at least one of these ideals is contained in $\cal I.$ Completely prime ideal is a two-sided ideal $\cal I\neq \cal A$ such that for two elements $x, y$ of the algebra $xy\in \cal I$ implies that at least one of these elements belongs to $\cal I.$ For commutative algebras these two notions coincide, prime ideals of the algebras of polynomials correspond to subvarieties. In general every completely prime ideal is prime, but prime ideal is not necessarily completely prime. However, in algebras we are interested in these two notions coincide. 

Every commutative  ring $\cal R$ without zero divisors (entire ring) can be embedded in the so called field of fractions $\cal F$  (formal expressions of the form $ab^{-1}$ with some identifications and natural operations). It is easy to check that any homomorphism of $R$ into a field has a unique extension to $F$. For non-commutative rings without zero divisors we can take this property as a definition. We say that a skew field $\cal {F}$ is a skew field of fractions of the ring $\cal A$ if $\cal A$ is a subring of $\cal F$ and every homomorphism of $\cal A$ into a skew field can be extended in unique way to a homomorphism of $\cal F.$  (Skew field is a non-commutative ring, where every non-zero element is invertible.)
 It is not always possible to construct the skew field of fractions, however, in many interesting situations it does exist.
For example, it is sufficient to assume that $\cal  A$ has no zero divisors and has at most polynomial growth, i.e. its Gelfand-Kirillov dimension (see \cite{Artin} for definition ) is finite (\cite{Berenstein} Lemma A1). In particular, one can consider the fraction skew  field of non-commutative torus $T$ defined as
the algebra  with generators $x_i,x^{-1}_i,1\leq i\leq n$ satisfying $x_jx_i=q_{ij}x_ix_j, x_ix^{-1}_i=x^{-1}_ix_i=1).$

 \section {Cluster algebras}
Let us consider an associative unital algebra $\mathcal {A}$ with generators $\{A_i\}$. We denote the set of generators $\{A_i\}$ by $\cal S;$ we use the same notation for the corresponding set of indices.  Let us fix a subset $  {F}\subset \cal{S}$ . We assume that the generators in $ {F}$  quasi-commute, i.e. 
$$A_iA_l={q _{il}}A_lA_i.$$   The set  $ {F}$ is called the set of frozen variables. We say that a finite set $K\subset \cal S$ is a cluster if  the generators $A_i\in K $ quasi-commute,
$K\supset  {F}$ and all other quasi-commuting systems of generators containing $ {F}$ have cardinality $\leq |K|.$ We denote $|K|$ (the number of elements in a cluster) by $k.$ 
Algebra  generated by $A_i\in K $ will be denoted by $\bf K.$ 

  We can consider an extension $\tilde {\cal A}$ of the algebra $\cal A$ where all generators $A_i$ are invertible elements.
  Then for every cluster $K=\{A_{i_1},...,A_{i_k}\}$ we can consider a subalgebra of   $\tilde {\cal A}$ generated by   
  $A_{i_1},...,A_{i_k},A_{i_1}^{-1},...,A_{i_k}^{-1}$ (a non-commutative torus denoted by ${ T}$).

Let us us suppose that the subset $L$ of cluster $K_1$  contains $k-1$ elements, hence one can obtain $K_1$  from $L$ adding a generator $A_n$. We assume that $L\supset F.$ We will describe how  one can obtain a new  cluster $K_2$  from $L$ adding a generator $A_m$  (in other words $K_1=L\cup \{A_n\}, K_2=L\cup \{A_m\}).$ We will consider monomials with respect to $L$, i.e. expressions of the form $ r\prod A_i^{\alpha^i}$ where $r\in \mathbb{C}.$ (We assume that the product as well as the products below run over $L.$ We should fix the order of factors in some way,  the change of the order  can be absorbed into the change of the constant factor $r$.)  We say that the transition from $K_1$ to $K_2$ is a mutation (more precisely an A-mutation) in the direction $A_n$ if $A_mA_n$ is a sum of two  monomials with respect to elements of $L$
  \begin {equation}
  \label {mu}
  A_mA_n=r\prod A_i^{\alpha^i}+s\prod A_i^{\beta^i}=rV+sU
  \end {equation}
and the following conditions are satisfied

a) $\alpha^i, \beta^i$ are non-negative integers obeying  $\alpha^i \beta^i=0$  (i.e. each factor $A_i$ enters only in one of the monomials),

b) For  $j\in K_1, j\neq n$  (i.e. $j\in L$)
\begin{equation}\label{E:qeq}
\begin{split}
& \prod {q} _{ji}^{\alpha^i}= \prod {q}_{ji}^{\beta_i},\text{ but }\\
& \prod {q} _{ni}^{\alpha^i}\neq \prod {q}_{ni}^{\beta^i}.
\end{split}
\end{equation}
It is easy to check that these conditions guarantee that  the element $A_m$ defined by the formula (\ref {mu}) in terms of the elements of the cluster $K_1$ quasicommutes with $A_i, i\in L$. If $A_m$ is a generator of $ \cal {A}$ we get a new cluster $K_2$.

We will be writing   $B=VU^{-1}$ for an element $\prod A_i^{\alpha_i}({\prod A_i^{\beta_i}})^{-1}\in {T}_1$  in the noncommutative torus . We can express conditions (\ref{E:qeq}) saying that $B$ commutes with all $A_i\in L$, but does not commute with $A_n$ (i.e. $B$ does not belong to the center of the non-commutative torus ${{T}}_1$).
%In formulas above all  products are over  $L$ (considered as a set of generators or as a corresponding set of indices).
 
 If  the clusters $K_1$ and $K_2$ are related by mutation then they are embedded in the same  non-commutative torus: $T_1$ = ${T}_2$  .  We will embed this torus into its skew field of fractions denoted by $\cal F$. There exists a unique homomorphism  $\phi_B$ of the torus   into $\cal F$ obeying $\phi_B(A_i)=A_i$ for $A_i\in L$, $\phi_B(A_n)=A_n(r+sB)^{-1}$, $\phi_B(A_i^{-1})=A_i^{-1}$ for $A_i\in L$, $\phi_B(A_n^{-1})=(r+sB)A_n^{-1}$. (The homomorphism is unique, because it is defined on generators of ${T}_1$, it exists because the above formulas preserve the commutation relations between generators.)  It can be extended in unique way to an automorphism of $\cal F$ denoted by the same symbol. 
 
 It is easy to check that 
 \begin{equation}
\label{a}
 \phi_B(UA_n^{-1})=(rV+sU)A_n^{-1}=A_m.
\end{equation}

We say that $\cal A$ is a poor man cluster algebra if

a) Every quasi-commuting family of generators containing $F$  belongs to a cluster;

b) One can get any cluster from any other cluster by means of a sequence of mutations.

Let us assume that the the defining relations of the algebra $\cal A$ depend on parameter $q$ (e.g. $q_{ij}$ are powers of $q$) in such a way that for $q=1$ we obtain a commutative algebra. It is convenient to assume that as we vary $q$ linear spaces $\mathcal{A}_a:=\mathcal{A}|_{q=a}$ form a continuous family. Technically it means that $\cal A$ is flat over $\C[q,q^{-1}]$.  Under certain conditions $\mathcal{A}_1$  can be identified with the algebra of polynomial homogeneous functions on a projective variety $P$. Then we write the quasi-commutation relations in the form
$A_jA_l=q^{\lambda _{jl}}A_lA_j$
where $\lambda _{jl}=-\lambda_{lj}\in \mathbb{Z}$
and the mutation should have the form
  $$A_mA_n=rq^a\prod A_i^{\alpha_i}+sq^b\prod A_i^{\beta_i}.$$
 The conditions imposed on the monomials in this formula will be specified later ( formula (\ref {m})). 
  It is convenient  to work  in non-commutative torus generated by   
  $A_{i_1},...,A_{i_k},A_{i_1}^{-1},...,A_{i_k}^{-1}$. We will be writing   $M( \rho), \rho =(\rho^1 ,\dots,\rho^k )\in {\mathbb{Z}}^k$ for the elements

  $$ M(\rho)=q^{\frac{1}{2}\sum_{l<i}\lambda_{il}\rho^i\rho^l}A_{i_1}^{\rho^1}\cdots A_{i_k}^{\rho^k}.$$
These elements coincide with $A_{i_j}$ when $\rho=e_j$ is an element of the standard basis for $ \mathbb{Z}^k$ and satisfy the following relations
  $$ M(\rho) M(\sigma) =q^{\frac{1}{2} \Lambda (\rho,\sigma)} M(\rho +\sigma),$$
  ( Here $ \Lambda (\rho,\sigma)=\sum_{j,l=1}^{k}\lambda_{jl}\rho^j\sigma^l.$)
  
  In these notations
  $$ A_m={\rm const} M(-e_n+{\bf {\alpha} })+{\rm const} M(-e_n+{\bf {\beta}})$$
  where ${\bf {\alpha}}=\sum _{i=1}^{k}\alpha^ie_i, {\bf {\beta}}=\sum _{i=1}^{k}\beta^ie_i, \alpha^n=\beta^n=0, \alpha^i\geq 0,\beta^i\geq 0$ and for every $i$ either $\alpha^i$ or $\beta^i$ vanishes.
  
  We introduce the notation  $b={\bf {\alpha}}-{\bf{\beta}}$ and impose the conditions
  \begin{equation}
\label{m}
\Lambda (b, e_i)=0\\ {\text{  for } }\\ i\neq n, \\
\Lambda (b,e_n)=-d(b).
\end{equation}
Here $d(b)$ denotes the minimal positive value for $\Lambda (b,\rho)$ where $\rho$ runs over ${\mathbb{Z}}^k.$

In the construction of the automorphism  $\phi_B:{\cal{F}}\to {\cal{F}}$ we use the same formulas with $B=M(b).$

  Notice that in the limit $q\to 1$ we obtain Poisson structure on the variety $P$ and on  classical limits of clusters, these Poisson structures agree.( The embedding of the algebra generated by cluster into the algebra $\cal A$ induces a Poisson map.)  
  
 A quantum cluster algebra  in the sense of \cite {BZ}  can be described as poor man cluster algebra where a mutation exists in the direction of any non-frozen element of the cluster.  %To specify such an algebra  one should fix a cluster and find a matrix $B$ of the size $k\times (k-f)$ having integer entries and obeying the relation $B^T E=D$  where $D$ is a diagonal matrix. (Here $f$ stands for the number of frozen variables and $E$ denotes the matrix appearing in quasi-commutation relations.) The matrix $B$ governs the mutations, the requirement that the mutation  in the direction of the element $A_n$ agrees with the mutation described above specifies one of the columns of the matrix $B$: ....
 In the construction of such an algebra Berenstein and Zelevinsky start with integer-valued skew-symmetric form $\Lambda$ 
 on  ${\mathbb{Z}}^k.$ Then they assume that there exists a solution $b_j\in {\mathbb{Z}}^k$ 
 of the equation
 $$ \Lambda (b_j,e_l)=-\delta_{lj }d(b_l)$$
 for every   non-frozen variable. (As earlier $d(b)$ denotes the minimal positive value for $\Lambda (b,\rho)$ where $\rho$ runs over ${\mathbb{Z}}^k.$) This allows them
  to define a mutation in  every non-frozen direction $e_j$ using the formula
 $$ A_m={\rm const} M(-e_j+{\bf {\alpha} }_j)+{\rm const} M(-e_j+{\bf {\beta}}_j)$$
 where ${\bf {\alpha} }_j$ stands for the positive part of $b_j$ and  ${\bf {\beta}}_j$ stands for its negative part. 
 
  Let us consider  prime ideals in $\cal A$ that are generated by  subsets of the set $\cal S$ (set of generators of $\cal A$). For every ideal  we consider the subset of $\cal S$ belonging to this ideal; such subsets will be called ideal subsets (of course, an ideal subset generates the corresponding ideal, but the ideal can be generated also by smaller subset). One can hope that taking a quotient of a cluster algebra with respect to the ideal of this kind we obtain a cluster algebra.

\section{Quantum spaces}
We have introduced the notion of non-commutative torus $T$ as an algebra generated by $x_i,x^{-1}_i,1\leq i\leq n$ which satisfy $x_jx_i=q_{ij}x_ix_j, x_ix^{-1}_i=x^{-1}_ix_i=1.$We say that an algebra $T$ is a based non-commutative torus if the system of generators is fixed. An isomorphism of two non-commutative tori is called base change if it transforms  generators of one torus into monomials of generators of another torus:  ${\tilde x}_i= c_ix_1^{a_{i1}}...x_n^{ a_{in}}$ where $a_{ij}$ is an invertible matrix over $\mathbb{Z} $, i.e. both the matrix and its inverse have integer entries.(One can prove that all isomorphisms of non-commutative tori have this form.)  Every isomorphism of non-commutative tori induces an isomorphism of their skew- fields of fractions, hence we  can talk about base change in these skew  fields. (Keep in mind that skew  fields have also other isomorphisms.)

Let us consider a family of non-commutative based tori $T_{\bf i}$ and their skew  fields of fractions
$\mathcal{F}_{\bf i},$  labeled by index $\bf i.$ We define an A-automorphism of a skew  field corresponding to a torus with generators $x_i,x^{-1}_i,1\leq i\leq k$ by the formula   $\phi_B(x_i)=x_i$ for $i\neq n$, $\phi_B(x_n)=x_n(r+sB)^{-1}$, $\phi_B(x_i^{-1})=x_i^{-1}$ for $i\neq n$, $\phi_B(x_n^{-1})=(r+sB)x_n^{-1}$ where $B$ is a monomial that does not contain $x_n, x_n^{-1}$. We  say that an isomorphism $\mathcal{F}_{\bf i}\to \mathcal{F}_{\bf i'}$ is an A-mutation if it can be represented as a composition of an A-automorphism  and base change.
The family is called  (a poor man) quantum A-space if every two elements of it can be connected by a sequence of A-mutations. 

Every poor man cluster algebra ( hence every quantum cluster algebra in the sense of \cite{BZ})  generates a  quantum A-space: tori $T_{\bf i}$ correspond to clusters. The  representation  of mutation of clusters as a composition of an A-automorphism and base change follows immediately from (\ref {a}).

We say that automorphism $\rho$ of $F_{\bf i}$ is an X-automorphism  if it leaves one of the generators (say $x_n$) intact and multiplies other generators by a polynomial of $x_n$ or by an inverse of such a polynomial: $\rho (x_n)=x_n, \rho (x_j)=x_j P_j(x_n)$ or $ \rho (x_j)=x_j {Q_j(x_n)}^{-1}$  for $j\neq n.$  A composition of an X-automorphism with base change is called X-mutation (in the direction $n$). By definition, a  family of based tori $T_{\bf i}$ and their skew  fields of fractions
$F_{\bf i} $ is a poor man quantum X-space if every two fields of fractions in this family are connected   by a sequence of mutations.

The definitions above are  downgraded versions  of the definitions in \cite {FG},\cite{FGQD} : we singled out only the properties of quantum spaces that are necessary for our proofs.   Fock and Goncharov  gave a construction of quantum spaces starting with some algebraic data. They \cite{FGQD} start with "feeds" $\mathbf{i} = (I, \epsilon_{ij} , d_i)$ where $I$ is a finite set, $d_i \in \mathbb{Q}_{>0}$,
and $\epsilon_{ij}$ , $i, j\in I$, is an integral valued matrix such that $d_i\epsilon_{ij}$ is skew-symmetric. Matrix $\hat{\epsilon}_{ik}:=d_i\epsilon_{ik}$ defines a skew-symmetric pairing $\Lambda:\Z^{I}\times \Z^{I}\to \frac{1}{N}\Z$, where $N$ is the least common multiple of the $\{d_i\}$. We introduce a notation $q_k:=q^{1/d_k}\in \Z[q^{1/N},q^{-1/N}]$
An X-automorphism $\nu_k$ of a  fraction field $F$ of a torus with generators $x_1, ...,x_n$ can be defined by the formula :
\begin{equation}\label{E:nudef}
x_i\to
\begin{cases}
&x_i(1+q_kx_k)(1+q_k^3x_k)\cdots(1+q_k^{2|\epsilon_{ik}|-1}x_k)\\
&x_i\left((1+q^{-1}_kx_k)(1+q_k^{-3}x_k)\cdots(1+q_k^{1-2|\epsilon_{ik}|}x_k)\right)^{-1}\\
\end{cases}
\end{equation}
( formula (60) from \cite{FGQD} ). The first line of this formula corresponds to $\epsilon_{ik}>0$, the second line corresponds to $\epsilon_{ik}<0$.  The X-mutation is a composition of X-automorphism  with base change.

The corresponding quantum X-space can be obtained from the initial feed by means of consecutive  application of X-mutations. The same initial data can be used also to construct a quantum A-space.

\section{Quantum Grassmannian}

The quantum Grassmannian $\Bbb{C}_q [ \Bbb{G}_{m,n} ]$ can be defined as an algebra with generators $\Delta _I$ where $I$ is an ordered subset of the set $[n]=\{1, 2, ...,n\}$ consisting of $m$ elements. ( Hence the set of generators $ \cal S$ can be identified with  $\{I\subset [n]||I|=m\}$).These generators (quantum minors)
satisfy quantum Pl\"ucker relations
\[ \sum_{i \in I-J} \ (-q)^{\inv(i,I) - \inv(i,J)} \ \Delta_{I-\{i\}} \ \Delta_{J \sqcup \{i\}} \ = \ 0 \]
\\

%The quantum Grassmannian $\Bbb{C}_q [ \Bbb{G}_{k,n} ]$, as defined in [10], is the $\Bbb{C}(q)$-algebra with unity
%generated by all quantum Pl\"ucker coordinates
%$\Delta^K$ where $K$ is a $k$-subset of $[1 \dots n]$ subject to the relations:

%\[ \sum_{i \in I-J} \ (-q)^{\inv(i,I) - \inv(i,J)} \ \Delta^{I-\{i\}} \ \Delta^{J \sqcup \{i\}} \ = \ 0 \]
%\\
\noindent
for any $(m+1)$-subset $I$ and $(m-1)$-subset $J$. Here $\inv(i,X)$ is the number of $x \in X\subset [1,\dots,n]$
such that $i>x$.

In the limit $q\to 1$  quantum Pl\"ucker relations give conventional Pl\"ucker relations. Hence the underlying projective variety $P$ coincides with the classical Grassmannian in this case.

Two  minors $\Delta_{I_1}, \Delta_{I_2}$  are quasi-commuting iff  $I_1=R\sqcup A_1, I_2=R\sqcup A_2$ where $A_1, A_2$  are cyclically separated (i.e. after some cyclic shift they belong to non-overlapping intervals). This condition  (called weak separation)  can be represented as a requirement that either $I_1=R\sqcup B\sqcup C, I_2=R\sqcup A_2, B<A_2<C$ or $I_1=R\sqcup A_1, I_2=R\sqcup D\sqcup E, D<A_1<E.$ In these notations
$\Delta_{I_1} \Delta_{I_2}= q^{\epsilon _{12}}\Delta_{I_2} \Delta_{I_1}$ where $\epsilon _{12}= |C|-|B|$ in the first case and $\epsilon _{12}=|D|-|E|$ in the second case \cite {SCO}.
%\[ \Delta^{I_{ij}} \ \Delta^{I_{st}} \ = \ \Delta^{I_{is}} \ \Delta^{I_{jt}} \ + \ \Delta^{I_{it}} \ \Delta^{I_{sj}} \] 

A  maximal collection of quasi-commuting minors ( a cluster) consists of $m(n-k)+1$ elements. (We assume that the set $\cal F$ of frozen variables is empty.)  Every collection of quasi-commuting minors can be extended to a maximal collection \cite {OPS}.

In the case when $m=2, n=4$ we have two clusters $(1,2),(2,3),(3,4),(4,1), (1,3)$ and $(1,2),(2,3),(3,4),(4,1),(2,4)$ The mutation can be written in the form

$\Delta_{13}\Delta_{24}=q\Delta_{12} \Delta_{34} + q^{-1}\Delta_{14} \Delta_{23}= 0$

Let us consider the general case. We will use the notation  $Rab$ for a union of some set   $R\subset [n]$ and two-element set $\{a,b\}$.  Let us assume now that a cluster contains minors with  $I=Rab, Rbc, Rcd, Rda, Rac$ where $a,b,c,d$ are cyclically ordered elements of $[n]\setminus R.$  Then we can get a new cluster  replacing the minor $\Delta_{Rac}$ with the minor $\Delta_{Rbd}$. The new cluster is related with the old one by mutation

$$\Delta_{Rac}\Delta_{Rbd}=q\Delta_{Rab}\Delta_{Rcd}+q^{-1}\Delta_{Rad}\Delta_{Rbc}$$

It is easy to check that the conditions (\ref {m}) are satisfied.

Every two clusters are connected by a sequence of mutations of this kind \cite {OPS}, hence 
$\Bbb{C}_q [ \Bbb{G}_{m,n} ]$ can be considered as poor man cluster algebra.

 For generic $q$ the prime ideals generated by a subset of the set of generators of quantum Grassmannian were classified in  \cite {LLR}.  (This classification was stated there as a conjecture, the proof was given in  \cite {LLN}.)  Recall that for every ideal of this kind we define  the ideal set as the subset of $\cal S$ consisting of generators belonging to this ideal. It was shown in  \cite {LLR} that   for quantum Grassmannian  the ideal sets are complements to positroid subsets of ${\cal {S}}=\binom {[n]}{m}$  defined by Postnikov \cite {P}. (Positroid subsets correspond to the cells of positive Grassmannians - positroid cells.)This means that the algebra obtained from  $\Bbb{C}_q [ \Bbb{G}_{m,n} ]$ by means of factorization with respect to  prime ideal generated by a subset of  family of minors quantizes the algebra of functions on a positroid cell. (All these ideals are completely prime, hence the quotient algebra is an entire ring.)

The positroid subsets (as well as prime ideals ) can be labeled by decorated permutations (permutations where every fixed point is decorated by an element of a two-point set). For a   decorated permutation $\sigma$ we denote the corresponding prime ideal by $\cal {I}(\sigma)$, the corresponding ideal set by $\{{\cal {I(\sigma)}}\}$ and the positroid $\cal S\setminus \{{\cal {I(\sigma)}}\}$ by $Pos(\sigma)$  The dimension of corresponding positroid cell is  denoted by $l(\sigma)$.
It is natural to assume that   both the  the quotient algebra  $\Bbb{C}_q [ \Bbb{G}_{m,n} ]/\cal {I}(\sigma)$ and the subalgebra of  quantum Grassmannian generated by $Pos(\sigma)$ are cluster algebras  with the size of cluster equal to $l(\sigma)+1$. The first of these statements was not proved yet.  The second one is proved under assumption that the set of frozen variables is defined as Grassmann necklace  corresponding to the permutation $\sigma$ \cite{OPS}.(One can prove also that there exists a family of minors  $Inn(\sigma)\subset Pos (\sigma)$  generating a cluster algebra with the same clusters and empty $\cal F$ \cite {KO}.)

\section {Hochschild homology and cohomology}

In the section we give some basic definitions and formulate some results about Hochschild homology and cohomology of non-commutative tori. The proofs are relegated to Appendix.

The standard  complex   used for definition of  Hochschild homology $HH_{*}(A)$ (see e.g. \cite{Loday}) is a direct sum of $C_n=A\otimes A^{\otimes n}, n\geq 0$. It is common practice to denote 
an element $a_0\otimes a_1\otimes \cdots \otimes a_n\in C_n$ by $a_0|a_1|\cdots|a_n.$
The differential is defined by the formula
\[\sd a_0|a_1|\cdots|a_n=a_0a_1|\cdots|a_n +\sum_{i=1}^{n-1} (-1)^{i}a_0|a_1|\cdots |a_ia_{i+1}|\cdots |a_n+(-1)^na_na_0|a_1|\cdots|a_{n-1}\]

Similarly one can define Hochschild homology with coefficients in bimodule 
$M$ (in this case the differential acts on the direct sum of $M\otimes A^{\otimes n}, n\geq 0$).

The Hochschild cohomology of the algebra $A$ with coefficients in bimodule $M$ can be defined by means of a complex represented as a direct sum of $C^n(A,M)$  where elements of $C^n(A,M)$ (cochains) are $n$-linear functions on $A$ with values in $M.$
The differential $d$ on the chain $D$ is defined by the formula
\[(dD)(a_0,\dots,a_n)=a_0D(a_1,\dots,a_n)+\sum_{i=0}^{n-1} (-1)^{i+1}D(a_0,\dots,a_ia_{i+1}\dots,a_n)+(-1)^{n+1}D(a_0,\dots,a_n)a_n\]

In particular, one-dimensional Hochschild cocycles are derivations, i.e. linear  maps $D:A\to M$ obeying  $D(ab)=D(a)b+aD(b).$

Notice that  in the case when $M=A$ the multiplication of cochains induces graded commutative multiplication of Hochschild cohomology classes \cite{Gerst}.

\begin{proposition}
We assume that associative algebra $A$ admits a skew  field of fractions $\mathcal{F}$
There is an action of cohomology on homology 
\begin{equation}\label{E:pairing}
H^i(A,A)\cap H_j(A,A)\to H_{j-i}(A,A),\quad  H^i(\mathcal{F},\mathcal{F})\cap H_j(\mathcal{F},\mathcal{F})\to H_{j-i}(\mathcal{F},\mathcal{F}).
\end{equation}
\end{proposition}
See \cite {TT} for the analysis of this action that generalizes the pairing of polyvector fields and differential forms with values in differential forms.
It can be described in the following way. Fix $D\in \Hom(A^{\otimes i},A)$ a Hochschild cochain. Then on the level of chains the pairing (\ref{E:pairing}) up to a sign is given by the formula (\cite{T}, formula 2.32)
\begin{equation}\label{E:cap}
D\otimes a_0|a_1|\cdots|a_j\overset{\cap}{\to} a_0D(a_1,\dots,a_i)|a_{i+1}|\cdots|a_j
\end{equation}

For smooth commutative algebras Hochschild homology classes can be identified with differential forms and Hochschild cohomology classes with polyvector fields.

Hochschild homology and cohomology of an   algebra with quasi-commuting generators were calculated in \cite{GG}.  One can use the  considerations of \cite {GG} to construct some special elements of the Hochschild homology of non-commutative torus, i.e.  of  the algebra $A$  generated by $x_i,x^{-1}_i,1\leq i\leq k$ which satisfy $x_jx_i=q_{ij}x_ix_j, x_ix^{-1}_i=x^{-1}_ix_i=1$. {\it (Later in this section $A$ stands for this algebra.)}
We will give a direct construction of these elements.

\begin{proposition}\label{1}
We will be writing $c_{\sigma(1),\dots,\sigma(k)},\sigma\in S_k$ for $\left(x_{\sigma(1)}\cdots x_{\sigma(n)}\right)^{-1}|x_{\sigma(1)}|\cdots |x_{\sigma(k)}$. Here $S_k$ stands for symmetric group on $k$ letters.The element
\[c=1/k!\sum (-1)^{|\sigma|}c_{\sigma(1),\dots,\sigma(k)}\]
is a Hochschild cycle. (One can say that this cycle is obtained from $\left(x_{\sigma(1)}\cdots x_{\sigma(n)}\right)^{-1}|x_{\sigma(1)}|\cdots |x_{\sigma(k)}$ by means on antisymmetrization.)
Its homology class is denoted by $[c].$
\end{proposition}

This statement can be derived from \cite {GG} (see Appendix for the derivation) or proved by direct calculation.

Notice that in the case of commutative torus the class $[c]$ can be identified with the standard volume form $\omega.$ In other words $[c]$ should be considered as $q$-deformed volume form.

\begin{remark}\label{R:cycles}
There is a simple way to construct more cycles. Fix a subalgebra $B\subset A$ isomorphic to a quantum torus with generators $x'_1, \dots,x'_{n'}$. Then the image of the antisymmetrization of $\left(x'_{\sigma(1)}\cdots x'_{\sigma(n')}\right)^{-1}|x'_{\sigma(1)}|\cdots |x'_{\sigma(n')}$ in $HH_{*}(A)$ will be a nontrivial homological class. Applying this construction to subalgebras generated by a subset  of generators of $A$ we get a family of cycles $c_R$ labelled by subsets $R\subset [k]=\{1,\ldots,k\}$.  It follows from the results of \cite{GG} that every class of the Hochschild homology  $HH_{*}(A)$ contains precisely one cycle represented as a linear combination of cycles $c_R$ with coefficient from the center of the algebra $A.$
\end{remark}
Hochschild cohomology and homology of   $A$ are related:

\begin{proposition} \label {pc}
%On a $A$-bimodule $M$ we define a new bimodule structure $M_f$ by the formula $x_i\times a\times x_j:=fx_if^{-1} a x_j$, $f:=x_1\cdots x_n $. 
$HH^i(A,A)\cong HH_{k-i}(A,A)$
%\begin{enumerate}
%\item $HH^i(A,M_f)\cong HH_{n-i}(A,M)$ and 
%\item $HH^i(A,A^{e})=0$ if $i\neq n$ and $HH^n(A,A^{e})=A_{f^{-1}}$
%\end{enumerate}
\end{proposition}
\begin{proof}
See Corollary \ref{C:one-dim}.
\end{proof}    
If $q_{ij}$ are not roots of unity non-commutative torus $A$ can be embedded in its field of fractions $\mathcal{F}.$

\begin{proposition}\label{P:restrictioniso}
Inclusion $A\to \cF$ induces  isomorphisms $H_*(A,\mathcal{F})\cong H_*(\mathcal{F},\mathcal{F})$, $H^*(A,\mathcal{F})\cong H^*(\mathcal{F},\mathcal{F})$.
\end{proposition}
See  Appendix for the proof.

 Let us consider derivations   $D_i$ obeying $D_ix_i=x_i, D_ix_j=0$ for $j\neq i$. ( Recall that $x_i$ are generators of the torus $A.$) Their cohomology classes will be denoted by $[D_i]\in H^1(A,A)$. We use cup-product in Hochschild cohomology to define an element
\[[\mu]=[D_1]\cup\cdots\cup [D_n]\in H^n(A,A)\] represented by cocycle $\mu=D_1\cup\cdots\cup D_n.$ (More explicitly, $\mu (\xi_1,\dots,\xi_n)=D_1(\xi_1)\dots D_n(\xi_n).$)
The derivations $D_i$ can be extended to the skew  field of fractions $\cF$ therefore we can use the same formula to define a cocycle $\mu$ and an element $[\mu]\in H^n(\mathcal{F},\mathcal{F}).$

%\begin {conjecture}
%$H^*(\mathcal{F},\mathcal{F})$ is a Grassmann algebra with $n$ generators $[D_1],\cdots, [D_n]$ over the center of $\mathcal{F}$. In particular, $H^n(\mathcal{F},\mathcal{F})$ is a one-dimensional vector space over the center of $\mathcal{F}.$
%\end{conjecture}

%We managed to prove a weaker statement (Proposition \ref{P:22}).
\begin{proposition}\label{ff}
$H_n(\mathcal{F},\mathcal{F})$ is a one-dimensional vector space over the center of $\mathcal{F}.$
\end{proposition}
\begin{proof}
Follows from Corollary \ref{C:one-dim}.
\end{proof}

\begin{proposition}\label{6}
$[\mu]\neq 0\in H^n(\mathcal{F},\mathcal{F})$
\end{proposition}
This statement follows from 
\begin{proposition}\label{66}
$[c]\cap [\mu]=[1], [1]\neq [0]\in H_0(\cF,\cF)$
\end{proposition}
The first formula can be obtained by direct application of the formula  (\ref{E:cap}) for cap-product. The second follows from Lemma \ref{L:seventeen}.

\begin{proposition}\label{7}
Let us consider two elements  of the group $ H^n(\cF,\cF)$  represented by cocycles $\nu_1, \nu_2$ regarded as  polylinear  $\cF$-valued functionals on $\cF$.  Assume that the cocycles $A\nu_1, A\nu_2$ obtained from these functionals by means of anti-symmetrization coincide when the arguments are chosen as  generators $x_1, ...,x_n $ of $A$.  Then $[\nu_1]=[ \nu_2]$.
\end{proposition}
In this proposition we use the notation $(A\nu)(\xi_1, \dots,\xi_n)=1/n!\sum_{\sigma\in S_n} (-1)^{|\sigma|}\nu(\xi_{\sigma(1)},\dots,\xi_{\sigma(n)})$. The proof is based on Proposition \ref {ff} and on the remark that the anti-symmetrization does not change cohomology class.

%Recall that the cup-product is graded  symmetric hence anti-symmetrization of cocycle given by $n$-linear  functional gives a homologous cocycle.  Using this remark and the fact that $H^n(\cF,\cF)$ is a one-dimensional vector space over the center of $\cF$ we obtain the statement we need.

\section {Hochschild homology of quantum spaces}

We have defined quantum A-spaces and X-spaces as families of based tori $T_{\bf i}$ and their skew  fields of fractions
$\cF_{\bf i}$  connected by mutations.

 Proposition \ref {1}  gives a special element in Hochschild homology of  every torus $T_{\bf i}$ and therefore an element   $[c_{\bf i}]$ in the Hochschild homology of
$F_{\bf i}$ . Proposition \ref{6} gives a special element $[\mu_{\bf i}]$ in Hochschild cohomology.

\begin {theorem} \label{CC}
A-mutations and X-mutations preserve the  elements $[c_{\bf i}]$ and $[\mu_{\bf i}]$ up to a sign.
\end {theorem}
(This theorem is a "quantization" of the statements about behavior of volume elements with respect to  A- and X-mutation in \cite{A}, formula (15.18))
We start with a {\bf proof of this statement for} $\mu_{\bf i}. $ 

First of all we calculate the effect of base change on the derivations $D_i$. From the relation $D_i(x_1^{a_1}...x_n^{a_n})=a_i x_1^{a_1}...x_n^{a_n}$ we obtain that the base change transforms these derivations into linear combinations of them with coefficients given by  the matrix specifying the base change. To calculate the transformation of $[\mu]$ we use the fact that cup-product is graded symmetric. We see that this transformation is given by the multiplication by the determinant of the matrix  of base change.  This matrix has integer entries  and  is invertible over integers, hence the determinant of it is equal to $\pm 1.$ 

The next step is the consideration of the behavior of $\mu$ with respect to  A-automorphism $\phi=\phi_B$.  Again we start with action of this automorphism  on the derivations $D_i$. To calculate ${\tilde D}_j=\phi D_j\phi ^{-1}$ we notice that $\phi^{-1}(x_k)=x_k(r+sB), \phi^{-1}(x_i)=x_i$ for $i\neq k.$ We obtain ${\tilde D}_j(x_i)=0 $ for $j\neq i$, ${\tilde D}_i(x_i)=x_i,$
${\tilde D}_i(x_k)= x_k(r+sB)^{-1}sD_iB$, ${\tilde D}_k(x_k)=x_k.$
 (In these formulas $i\neq k$. )
  
 The most important  consequence of these relations is the remark that   ${\tilde D}_j(x_j)=D_j(x_j)$ for all $j$, hence ${\tilde \mu}(x_1, ..., x_n)=\mu (x_1,...,x_n).$ (Here $\tilde \mu$ stands for ${\tilde D}_1\cup\cdots\cup {\tilde D}_n.$) Notice that all terms that we add by anti-symmetrization vanish because  ${\tilde D}_j(x_i)=D_j(x_i)=0 $ for $j\neq i$. Therefore the relation $[{\tilde \mu}]=[\mu]$ follows from Proposition \ref {7}.
 
 The same proof works  for X-automorphism $\rho$. We define ${\tilde D}_j=\rho D_j\rho ^{-1}$ and check that  ${\tilde D}_j(x_j)=D_j(x_j)$ for all $j$  and  ${\tilde D}_j(x_i)=D_j(x_i)=0 $ for $j\neq i$. Again this allows us to apply Proposition \ref {7}.
 \paragraph{The proof that $c$ is invariant under mutations}
 To prove the invariance of $[c]$ with respect to A- and X-mutations we notice that $[c]\cap[\mu]=[1]$ (Proposition \ref {66}).
 It is a general fact that cap product is compatible with homomorphisms : $[\lambda]^{\psi}\cap [\tau]^{\psi}=([\lambda]\cap [\tau])^{\psi}$. It is clear that $[1]^{\psi}=[1]$ for any $\psi$. As we know  $[\mu]^{\phi}=[\mu]$ and $[c]\cap [\mu]=[1]$ then 
 \begin{equation}\label{E:cinv}
 [c]^{\phi^{-1}}\cap [\mu]=[c]\cap [\mu]^{\phi}=[c]\cap [\mu]=[1]
 \end{equation}
 Cup-product with $\mu$ defines a nonzero $Z(\cF)$-linear map from one-dimensional linear space $H_{k}(\cF,\cF)$ over center $Z(\cF)$ (Corollary \ref{C:one-dim}) to $H_{0}(\cF,\cF)$. One-dimensionality and (\ref{E:cinv}) imply that $[c]^{\phi}=[c]$
%   If we know $[\mu]$ this relation specifies $[c]$ uniquely (this fact is proven in the Appendix). Therefore the invariance of $[c]$ follows from the invariance of cap-product and $[\mu].$
 
 Notice that working in terms of power series with respect to $1-q$ one can represent the X-automorphism as a conjugation with quantum dilogarithm \cite{FGQD}. This allows us to simplify the proof in this setting (Hochschild (co)homology is invariant with respect to inner automorphisms).
 
 \section {Poisson and Hochschild homology}
 
 If a non-commutative  algebra ${\cal {A}}_q$ depends on parameter $q$ and becomes commutative for $q=1$ then the algebra ${\cal {A}}_1$
has a structure of Poisson algebra. We are interested in the case when ${\cal {A}}_1$
is an algebra of functions on a Poisson manifold (or an algebra of homogeneous functions on a projective Poisson variety). 

One can define Poisson homology $HP_i$ and cohomology $HP^i$ of Poisson algebra basically "taking the limit $q\to 1$" in the definition of Hochschild homology and cohomology ( a limit of bimodule over ${\cal {A}}_q$ as $q\to 1$ is a Poisson module over ${\cal {A}}_1$ hence one can talk  about Poisson homology and Poisson cohomology with coefficients in  Poisson module). %See for example \cite {POI} for definitions.)

The limit  of Hochschild homology class $c$ constructed in Sec 3 as $q\to 1$  is closely related to the differential form  constructed in \cite {FG} and used in \cite {A}. It follows that this form is a Poisson cycle. Of course, this can be checked directly.

It follows from the results of Section 3 that in the case when ${\cal {A}}_q$ is a non-commutative torus and $q$ is generic  the Hochschild homology $HH_i({\cal {A}}_q)$ 
is canonically isomorphic to Poisson homology  of commutative torus ${\cal {A}}_1.$ Similar results are valid for Hochschild and Poisson cohomology. ( Non-canonical isomorphism between Hochschild and Poisson is known in more general situations.) 
 The results of Section 4 imply in the limit $q\to 1$ that for two commutative tori related by mutation there exists a $q$-dependent family of isomorphisms of their Poisson homologies.

It was shown in \cite {FG} there there exists a canonical quantization of Poisson cluster algebras, it seems that using this result one can give an independent construction of the $q$-dependent family of isomorphisms we have mentioned.

\section {Contribution  of positroid cell to scattering amplitudes}

It was proved in \cite {A} that the contribution of positroid cell $\sigma \subset  G(k,n)$  to the amplitude can be described in momentum -twistor formalism
as an integral (8.27) containing the delta-function of ${\hat C} \cal Z$ where ${\hat C}\in {\hat \sigma}$, \ $\hat \sigma$ stands for the positroid cell in $G(k-2, n)$ and $\cal Z $ denotes a a column of $n$ supertwistors. (A supertwistor  is considered as  a pair $(z,\eta)$ where $z$ is momentum-twistor variable and $\eta$ is an odd variable; it was shown in \cite {AHT} that $\eta$ should be identified with $dz.$) We will give reformulation of this construction that allows us to quantize it. Moreover, this reformulation opens new ways for calculation of amplitudes.

To calculate the integral (8.27)  one should solve  the equation ${\hat C} {\cal Z}=0$ . It is sufficient to solve this equation for purely bosonic $\cal Z$ (the equation for the odd part has precisely the same form). If there exists a unique $\hat C$ obeying the equation for generic $\cal Z$ the solution specifies a map $\alpha$ of the top cell $ G(4,n+4)$ into the orthogonal positroid cell ${\hat \sigma}^{\bot}$ ( this top cell can be identified with the top cell of the space of $4\times n$  matrices). Then the amplitude is equal to $\alpha ^*\omega$ ( the pullback of the standard volume form $\omega$ on the positroid  ) for appropriate choice of $\alpha$. If the equation has finite number of solutions (the map $\alpha $ is multivalued),  then we can lift $\alpha $ to a single-valued map of some variety $M$ into ${\hat \sigma}^{\bot}$, the natural projection of $M$ onto $G(4,n+4)$ is a branched covering. ( The variety $M$ can be defined as the set of   pairs  ${\hat C}, {\cal Z}$ where $\cal Z$ is purely bosonic obeying  ${\hat C} {\cal Z}=0$.) Then 
we can construct the scattering amplitude taking first the pullback of the volume form $\omega$ to $M$ and then the push-forward to $G(4,n+4)$. (Denoting by $\alpha$ the projection of $M$ onto ${\hat \sigma}^{\bot}$ and by $\pi$ the projection onto  $G(4,n+4)$ we represent the differential form corresponding to the scattering amplitude as $\pi _!\alpha^*\omega.$)

The construction of scattering amplitude presented in this form can be easily quantized by means of the theorems we have proven. Instead of the positroid cell  ${\hat \sigma}^{\bot}$ we consider the corresponding noncommutative torus $T$, instead of top positroid cell in the  Grassmannian $G(4,n+4)$ we consider the noncommutative torus $U$ corresponding to this cell. 
If there exists a homomorphism $T \to U$ the $q$-deformed scattering amplitude can be defined as an image of $q$-deformed volume form (of Hochschild homology class $[c]$ constructed in Theorem \ref {1}). Moreover, it is easy to describe explicitly all homomorphisms of noncommutative tori; if the homomorphism we need does exist we get an explicit expression for  a single-valued map $\alpha$ in the classical limit. If such a homomorphism does not exist  we should consider intermediate noncommutative variety  $V$ that models the variety $M$ and construct homomorphisms $T\to V$, $U\to V$. All homomorphisms can be easily calculated.
Taking the limit $q\to 1$ we obtain an explicit expression for the contribution of the positroid cell into scattering amplitude in the situation of \cite {A}.
\vskip .1in

{\bf Acknowledgements} We are indebted to N. Arkani-Hamed, A. Berenstein, A. Goncharov, S. Launois,  T. Lenagan and J. Trnka for useful discussions.
 
 \section {Appendix}
 There is a remarkable complex associated with the algebra whose generators satisfy $y_iy_j=q_{ij}y_jy_i$ \cite{GG}, see also \cite {WA}. In fact the authors construct the complex in  greater generality.

Introduce  noncommutative algebras $X_*$ and $A$ over $\C[q]:=\C[q_{ij}]/(q_{ij}q_{ji}-1), 1\leq i,j\leq n$. $X_*$ is a quotient of a free associative algebra on generators $y_i,z_j,e_l,y^{-1}_i,z^{-1}_j 1\leq i,j,l\leq k$. The ideal is generated by the relations
\begin{equation}\label{E:relations0}
\begin{split}
&y_jy_i=q_{ij}y_iy_j, \quad z_jz_i=q_{ij}z_iz_j, \quad z_jy_i=q_{ij}y_iz_j\\
&y^{-1}_iy_i=y_iy^{-1}_i=z^{-1}_iz_i=z_iz^{-1}_i=1\\
\end{split}
\end{equation}
\begin{equation}\label{E:relations1}
\begin{split}
&e_je_i=-q_{ij}e_ie_j,\quad  e^2_i=0, \quad e_jy_i=q_{ij}y_ie_j, \quad e_jz_i=q_{ij}z_ie_j
\end{split}
\end{equation}
In positively graded algebra  $X_*$  $y_i, z_j$ have degree zero, $e_k$ has degree one.  The differential $\sd$ satisfies graded Leibniz rule and 
\[\sd y_i=\sd z_j=0, \sd e_l=y_l-z_l.\]

%We apply these construction to non-commutative torus 
Let $A$ be a noncommutative torus .
%based on generated by 
%$x_i,x^{-1}_i,1\leq i\leq n$ which satisfy 
%$x_jx_i=q_{ij}x_ix_j, x_ix^{-1}_i=x^{-1}_ix_i=1$.
$X_*$ is $A$-bimodule: $x_i\times a\times x_j:=y_iaz_j$ for any $a\in X_*$. Algebra  $A^{op}$ shares linear space with $A$ but has the product defined by the formula $a\times b:=ba$. Ant $A$-bimodule $M$ is the same thing as a left module over the algebra $A^{e}:=A\otimes A^{op}$. $(a\otimes b)m=amb$. It is also a right module over the same algebra $m(a\otimes b):=bma$.
$A$ is naturally an $A$-bimodule

\begin{proposition}
$X_*$ is a resolution of $A$ by free $A$-bimodules.  The augmentation map $\epsilon:X_*\to A$ sends $y_i,z_i$ to $x_i$ and $e_i$ to zero.
\end{proposition}
\begin{proof}
See \cite{GG}.
\end{proof}
%To emphasize that $X_*$ is a resolution of $A$ we will be writing $X_*(A)$.
We can use this complex for computation of Hochschild homology. To this end we take the tensor product  of $A$-bimodules $\XA:=X_*\underset{A^e}{\otimes} A$ and compute the cohomology of the resulting complex. Denote by $\Lambda_q$ the subalgebra in $X_*$ generated by $e_i$. 
The space $\XA$, which not an algebra any more, is spanned  by $a\otimes b, a\in \Lambda_q,b\in A$.  $\XA$ in addition to relations that come from (\ref{E:relations0},\ref{E:relations1}) has relation $a\otimes b=b\otimes a$. It comes from the tensor product of bimodules. 
%$\XA$ is a tensor product of bimodules over a noncommutative algebra and  we don't  expect  an algebra structure on this space that comes from a general construction.

Note that elements $e_iy^{-1}_i$ $e_iz^{-1}_i$ belong to the (graded) center of $X_*$
\begin{proposition}
Let $P\in X_*$ be a nonzero polynomial in $e_iy^{-1}_i$ The image of  $P\otimes 1$ defines a nonzero  $\sd$-cycle  in $\XA$. It
 also defines a nontrivial homology class in $HH_*(A)$. 
\end{proposition}
\begin{proof}
See \cite{GG}.
\end{proof}

%The standard  complex   used for definition of  Hochschild homology $HH_{*}(A)$ (see e.g. \cite{Loday}) is a direct sum of $C_n=A\otimes A^{\otimes n}, n\geq 0$. It is common practice to denote 
%an element $a_0\otimes a_1\otimes \cdots \otimes a_n\in C_n$ by $a_0|a_1|\cdots|a_n$
%The differential is defined by the formula
%\[\sd a_0|a_1|\cdots|a_n=a_0a_1|\cdots|a_n +\sum_{i=1}^{n-1} (-1)^{i}a_0|a_1|\cdots |a_ia_{i+1}|\cdots |a_n+(-1)^na_na_0|a_1|\cdots|a_{n-1}\]

The cycles corresponding to products of $e_iy_i^{-1}$ are described in the Proposition \ref{1}. Let us prove this Proposition.
\paragraph{Proof of Proposition \ref{1}}\label{Par:1}
\begin{proof}
According to  Proposition 1.4, Remark 1.5 \cite{GG} there exists a chain map $\theta_{*}:\XA\to C_{*}$.
\begin{equation}\label{E:cycle}\theta(\left(x_{1}\cdots x_{n}\right)^{-1}e_{1}\cdots e_{n})=\sum_{\sigma\in S_n} sg_q(\sigma)\left(x_{1}\cdots x_{n}\right)^{-1}|x_{\sigma(1)}|\cdots |x_{\sigma(n)}
\end{equation}
Here $sg_q(\sigma)=\prod_{h>j,\sigma(h)<\sigma(j)}(-q_{i_{\sigma(h)}i_{\sigma(h)}})=(-1)^{|\sigma|}\prod_{h>j,\sigma(h)<\sigma(j)}(q_{i_{\sigma(h)}i_{\sigma(h)}})=(-1)^{|\sigma|}\widetilde{sg}_q$. The factor $\widetilde{sg}_q$ can be absorbed by the element with negative exponent. We continue  equation (\ref{E:cycle}):
\[=\sum_{\sigma\in S_n} (-1)^{|\sigma|} \left(x_{\sigma(1)}\cdots x_{\sigma(n)}\right)^{-1}|x_{\sigma(1)}|\cdots |x_{\sigma(n)}\]
We obtained the cycle described in Proposition \ref {1}.
\end{proof}
The complex 
\begin{equation}\label{E:cohdef}
X^*(A):=\Hom_{A^e}(X_*,A)
\end{equation} computes Hochschild cohomology $HH^i(A,A)$.

The following two-term complex $P$ will be useful in our computations
\[\C[x,x^{-1}]\overset{x-1}\leftarrow \C[x,x^{-1}]\]
Its only nontrivial nonzero cohomology is in degree zero and is  to $\C$. We interpret $P^{\otimes n}$ as a projective resolution of $\C$ over $B=\C[x_1,x_1^{-1},\dots, x_k,\xi_k^{-1}]$. It is isomorphic to Koszul complex $K_*(B):=B\otimes\Lambda[\xi_1,\dots,x_n]$ with differential $\sum_{i=1}^n(x_i-1)\frac{\sd}{\sd \xi_i}$.  We have $K_i(B)=B\otimes\Lambda^i$. There is also a closely related complex $K^*(B):=B\otimes\Lambda[\xi_1,\dots,\xi_n]$ with graded components  $K^i(B)=B\otimes\Lambda^i$. We set $r=b\xi_{i_1}\cdots \xi_{i_s}$. Then $dr=\sum_{i=1}^k(x_i-1) \xi_ir$. The complexes are isomorphic under the map
$\psi:K^s(B)\to K_{n-s}(B)$
\[\psi (b\xi_{i_1}\cdots \xi_{i_s})=b\frac{\sd^s}{\sd\xi_{i_1}\cdots \sd\xi_{i_s}}\xi_{1}\cdots \xi_{n}.\]

$B$ is the group algebra of the group $\Z^k$. We can use $K_*$ and $K^*$ for computation of homology and cohomology of $\Z^n$:
\begin{equation}\label{E:duality}
\begin{split}
&H_i(\Z^k,M)=H_i(K(M))\\
&H^i(\Z^k,M)=H^i(Hom_B(K(B),M)=H^{i}K^*(M)=H_{k-i}K_*(M)
\end{split}
\end{equation}
It is convenient to choose $\eta_i=e_iy_i^{-1},y_i,z_i$ as a generating set for $X_*$.
The algebra $A$ is a projective representation of $\Z^n$: $x_jx_ia=q_{ij}x_ix_ja$. Because of that it wouldn't be a surprise to learn that in adjoint action $x_i\times a=x_iax_i^{-1}$ cocycle factors $q_{ij}$ drop out and we have an ordinary  (as opposed to projective) representation of $\Z^n$. We denote it by $A^{ad}$
\begin{proposition}\label{P:iso}
A map $a\eta_{i_1}\cdots \eta_{i_s}\to a\xi_{i_1}\cdots \xi_{i_s}$ defines an isomorphism of complexes $X_{*}(A)$ and $K_*(A^{ad})$.
Fix $g\in Hom_{A^e}(X_{*},A)$. We will be writing $g_{i_1,\dots,i_s}$ for the value of $g$ on $\eta_{i_1}\cdots \eta_{i_s}$.
The map $g\to \sum g_{i_1,\dots,i_s} \xi_{i_1}\cdots \xi_{i_s}$ defines an isomorphism of complexes $X^{*}(A)$ and $K^*(A^{ad})$. The last two statements are still valid if we replace $A^{ad}$ by $\cF^{ad}$.
\end{proposition}
\begin{proof}
Direct inspection.
\end{proof}
Denote by $\cF(x_1,\dots,x_k)$ the noncommutative field of fractions of $A(x_1,\dots,x_k)$.
\begin{corollary}\label{C:one-dim}
It follows from (\ref{E:duality}) and Proposition \ref{P:iso} that $HH_i(A,A)=HH^{k-i}(A,A)$, $HH_i(A,\cF)=HH^{k-i}(A,\cF)$. In particular $HH_{k}(A,\cF)=HH^{0}(A,\cF)$. Recall that $HH^{0}(A,\cF)=HH^{0}(\cF,\cF)$ is the center of $\cF$.
\end{corollary}

Fix a ring $B$ and    an injective automorphism $\phi$ of a $B$. Skew Laurent series   ring  $B((x, \phi))$ (see \cite{C} for details) consists of formal series  $b=\sum_{i=k}^{\infty}b_ix^i $ of an indeterminate $x$ with an integer $k$ (which could be negative) and coefficients $b_i \in  B$. Addition in $B((x, \phi))$ is defined naturally and multiplication is defined by the rule $x^ib = \phi^i(b)x (a \in B)$. The ring $B((x, \Id))=B((x))$ is the ordinary Laurent series ring of $B$.

According to \cite{C} Theorem 2.3.1 if $B$ is a skew field then $B((x, \phi))$ is also a skew field. We can construct this way examples of skew fields inductively. Start with $\C((x_1))$ and define an automorphism $\phi$ that acts by $\phi(x_1)=q^{-1}_{12}x_1$. The skew field $\C((x_1))((x_2,\phi))$ will contain subalgebra $A(x_1,x_2)$ generated by $x_1^{\pm},x_2^{\pm}$, which satisfy $x_2x_1=q_{12}x_1x_2$.  $\C((x_1))((x_2,\phi))$ will also contain the field of fractions $F(x_1,x_2)$. Suppose $\C((x_1))((x_2,\phi))\cdots ((x_{n-1},\phi))$ is given. Extend $\phi$ by the rule:$\phi(x_i)=q^{-1}_{1n}x_i$. The resulting skew field 
\begin{equation}\label{E:SFdef}
SF(x_1,\dots,x_n):=\C((x_1))((x_2,\phi))\cdots ((x_{n},\phi))
\end{equation} contains $A(x_1,\dots,x_n)$ and its field of fractions $\cF(x_1,\dots,x_n)$. 

\begin{proposition}\label{P:SFflatness}
$SF(x_1,\dots,x_n)$ is flat left (right) module over $A(x_1,\dots,x_n)$.
\end{proposition}
\begin{proof}
Let us carry the proof for the left structures.
$SF$ is a   linear space over $\cF$. It has a Hamel basis indexed by some set $Y$. Let  $Z$ be a set of finite subsets of $Y$ ordered by inclusion. Then $SF=\lim_{\underset{Z}\to}\cF^z, z\in Z$. We already know that $F$ is flat over $A$ and by Lazard, Govorov theorem $\cF=\lim_{\underset{Z}\to}A^z, z\in U$. From this we conclude that $SF$ is a direct limit of free $A$-modules indexed by the set $Z\times U$ and $SF$ is flat $A$-module.
\end{proof}

\begin{proposition}\label{P:resolution}
The tensor product $\cF^e\underset{A^e}{\otimes} X_{*}, (SF^e\underset{A^e}{\otimes} X_{*})$ is a free $\cF^e$ ($SF^e$) resolution of $\cF,(SF)$.
\end{proposition}
\begin{proof}
We do the proof only for $\cF$. The proof for $SF$ is similar. Flatness in this case follows from Proposition \ref{P:SFflatness}.

$\cF^e$ is $A^e$-flat. This is true because by Lazard, Govorov theorem $\cF=\lim_{\rightarrow} \cF_{\alpha}$ where $\cF_{\alpha}$ free finitely generated $A$-modules indexed by directed system. $\cF^{op}_{\beta}$ is a similar system of modules for $\cF^{op}$ over $A^{op}$. $\cF_{\alpha\beta}=\cF_{\alpha}\underset{\C}{\otimes}\cF^{op}_{\beta}$ is a bi-system over $A^{e}$ for $\cF^e$. $F_{\alpha\beta}$ are obviously $A^{e}$-free. Thus $\lim_{\rightarrow}\cF_{\alpha\beta}=\cF^e$ is flat.
\begin{lemma}\label{P:freemodule}
Any domain of polynomial growth is an Ore domain. In particular $A$ is an Ore domain
\end{lemma}
\begin{proof}
See \cite{BZ} Proposition 11.1
\end{proof}
It follows from flatness of $\cF^e$ and Lemma \ref{P:freemodule} that $\cF^e\underset{A^e}{\otimes} X_{*}$ is acyclic away from zero degree. Its cohomology at zero is $\cF^e\underset{A^e}{\otimes} A=\cF$. Indeed, by Ore condition  $\forall a,s\neq0 \in A,\exists a',s'\neq0 \in A$ $ s'\underset{A}{\otimes} a \underset{A}{\otimes} 1= 1\underset{A}{\otimes} s'a \underset{A}{\otimes} 1=1\underset{A}{\otimes} a's\underset{A}{\otimes} 1 =1\underset{A}{\otimes} a'\underset{A}{\otimes} s $. After multiplication on $s^{-1},s'^{-1}$ it becomes  $s'^{-1}\underset{A}{\otimes} a' \underset{A}{\otimes} 1=1\underset{A}{\otimes} a\underset{A}{\otimes} s^{-1}$. We use this to identify $\cF^e\underset{A^e}{\otimes} A$ with $\cF$.

\end{proof}
%\begin{proposition}
%The tensor product $F^e\underset{A^e}{\otimes} X_{*}$ is a differential graded algebra.
%\end{proposition}
%\begin{proof}

%\end{proof}
We do not expect that $\cF^e\underset{A^e}{\otimes} X_{*}$ will be an algebra for arbitrary $q$. The reason is that the product $ba$ $a\in \cF(y_1,\dots,y_n)\subset \cF(y_1,\dots,y_n,z_1,\dots,z_n)$ and $b\in \cF(z_1,\dots,z_n)\subset \cF(y_1,\dots,y_n,z_1,\dots,z_n)$ will in general lie in $\cF(y_1,\dots,y_n,z_1,\dots,z_n)$, which is much bigger then $\cF(y_1,\dots,y_n)\otimes \cF(z_1,\dots,z_n)$.

\paragraph{Proof of Proposition \ref{P:restrictioniso}}\label{Par:restrictioniso}
%\begin{proposition}\label{P:restrictioniso}
%The map $A\to \cF$ induces  isomorphisms $H_*(A,\cF)\cong H_*(\cF,\cF)$, $H^*(A,\cF)\cong H^*(\cF,\cF)$.
%\end{proposition}
\begin{proof}
By Proposition \ref{P:resolution} $\cF\underset{\cF^e}{\otimes}\cF^e\underset{A^e}{\otimes} X_{*}$ is the complex for calculation of $H_*(\cF,\cF)$. It coincides with $\cF\underset{A^e}{\otimes} X_{*}$ that computes $H_*(A,\cF)$. The argument for cohomology is similar.
\end{proof}
In the following we will use freely these isomorphisms.
\begin{proposition}
There is an action of cohomology on homology 
\begin{equation}\label{E:pairing}
H^i(A,A)\cap H_j(A,A)\to H_{j-i}(A,A),\quad  H^i(\cF,\cF)\cap H_j(\cF,\cF)\to H_{j-i}(\cF,\cF).
\end{equation}
\end{proposition}
\begin{proof}
The proof follows from two facts. The first is the  interpretation of $H^i(A,A)$ as $\Ext^i_{A^e}(A,A)$ and $H_i(A,A)$ as $\Tor_i^{A^e}(A,A)$ (see \cite{Weibel} Lemma 9.1.3). The second  is that  $\Ext^i$ is a group of morphisms  in derived category $\Hom_{DMod(A^e)}(A,A[i])$. It must define morphism of derived functors($A\overset{\mathbb{L}}{\underset{A^e}{\otimes} }?$ in our case)(see \cite{GM} for details). 

More explicitly the pairings have been described by \cite{TT} in the context of noncommutative calculus. Fix $D\in \Hom(A^{\otimes i},A)$ a Hochschild cochain. Then on the level of chains the pairing (\ref{E:pairing}) up to a sign is given by the formula (\cite{T} formula 2.32)
\begin{equation}\label{E:cap}
D\otimes a_0|a_1|\cdots|a_j\overset{\cap}{\to} a_0D(a_1,\dots,a_i)|a_{i+1}|\cdots|a_j
\end{equation}

\end{proof}
\begin{remark}\label{R:duality}
On a $A$-bimodule $M$ we define a new bimodule structure $M_f$ by the formula $x_i\times a\times x_j:=fx_if^{-1} a x_j$, $f=x_1\cdots x_n $. Then the complex $\{X^i(M_c),d\}$ is isomorphic to $\{X_{n-i}(M),d\}$. As a corollary we get  Proposition \ref {pc}.
\end{remark}

%We are interested in derivations defined by  $D_{x_i}=x_i\sd_i\in H^1(A,A)$. We can use cup-product in Hochschild cohomology to define an element
%\[\mu=D_{x_1}\cup\cdots\cup D_{x_n}\in H^n(A,A)\]
%\begin{proposition}\label{E:mu0}
%$\mu\neq 0\in H^n(A,A)$
%\end{proposition}
%\begin{proof}
%There is a quasi-isomorphism $\theta^*:C^*(A,A)\to X^*(A)$ (\cite{GG},Proposition 1.3) between the standard cohomological complex $C^i(A,A)=\Hom(A^{\otimes i},A)$ (see \cite{Loday} for details) and the complex (\ref{E:cohdef})
%By Remark 1.5 \cite{GG} the image  $\theta^*\mu$ in the complex $X^*(A)$ is equal to $\sum_{\sigma\in S_n}\mu(sg_{q}(\sigma)|x_{\sigma(1)}|\cdots|x_{\sigma(n)}|1)e_1\cdots e_n=x_1\cdots x_n e_1\cdots e_n$. It follows from item 2 Remark 1.5 \cite{GG} that $x_1\cdots x_n s e_1\cdots e_n$ is a trivial cocycle iff there are elements $a_i\in A$ such that $s=\sum_{i=1}^n[x_i,a_i]$. This is the condition that class $[s]$ is trivial in $H_0(A,A)$.
\begin{lemma}\label{L:seventeen}
$[1]\neq [0]\in H_0(A,SF)$
\end{lemma}
\begin{proof}
If $a_i$ are monomials $a_i=x^m_i:=x_1^{m_i^1}\cdots x_n^{m_i^n}$ then  $[x_i,a_i]$ will be a zero or non-constant   monomial . From this we conclude that equation 
\begin{equation}\label{E:trivialcocycleeq}
\sum_{i=1}^n[x_i,a_i]=1
\end{equation} has no solutions. 
Solution of this equation in $SF$ is the same as a sequence $a_i(r)\in A, r=1,\dots,\infty$, which converge in $SF$ to $a_i$ and such that $\sum_{i=1}^n[x_i,a_i(k)]-1$ converge to zero, that is divisible by arbitrary high power of $x_1$ in formal Taylor series in $x_1$. As  $-1$ will never cancel with any term in $\sum_{i=1}^n[x_i,a_i(k)]$ such divisibility is impossible(see \cite{C} for details on topology on $SF$).

%Thus $x_1\cdots x_n e_1\cdots e_n$ is a nontrivial cocycle.
\end{proof}
\begin{corollary}\label{C:zerohomology}
$[1]\neq [0]\in H_0(A,\cF)$ because $[1]$ is mapped to nonzero element in $ H_0(A,SF)$
\end{corollary}
\begin{proposition}\label{C:zerohomology2}
For generic $q_{ij}$ the center $Z(A)$ maps to $H_0(A,A)=A/[A,A]$ and the map is an isomorphism.
\end{proposition}
\begin{proof}
The proof will simplify if we use an action of a commutative torus $\prod_{i=1}^k\C^{\times}$ on $A$: $zx_i=z_ix_i$ where $z=(z_1,\dots,z_k)$. Weight spaces of $G^n$ action on $A$ are spanned by monomials $m$ in $x_i$. Weight grading of $A$ descends to $A/[A,A]$. We  have direct sum decomposition $A=\bigoplus_{m\in Z(A)} \C m+\bigoplus_{m\notin Z(A)} \C m=Z(A)+B$. If $m\notin Z(A)$, then there is $m'\notin Z(A)$ such that $mm'\neq m'm$. Still $mm'=F(q) m'm$, for some polynomial $F$ in $q_{ij}$. Thus $(m(m')^{-1})m'-m'm(m')^{-1})=m-F^{-1}(q)m$. We conclude that for generic $q_{ij}$ $B\subset [A,A]$. We conclude that the map $Z(A)\to A/[A,A]$ is onto. 

By Corollary \ref{C:zerohomology} $[1]$ maps to nonzero element in $\cF/[\cF,\cF]$. The later group is a linear space over the field $Z(\cF)$. By construction $Z(A)\subset Z(\cF)$. Thus $Z(A)\to A/[A,A]$ has no kernel.
\end{proof}
%\begin{remark}\label{R:duality}
%On a $A$-bimodule $M$ we define a new bimodule structure $M_c$ by the formula $x_i\times a\times x_j:=cx_ic^{-1} a x_j$, $c=x_1\cdots x_n $. Then the complex $\{X^i(M_c),d\}$ is isomorphic to $\{X_{n-i}(M),d\}$. As a corollary we get  that 
%\begin{enumerate}
%\item $HH^i(A,M_c)\cong HH_{n-i}(A,M)$ and 
%\item $HH^i(A,A^{e})=0$ if $i\neq n$ and $HH^n(A,A^{e})=A_{c^{-1}}$
%\end{enumerate}
%\end{remark}
Let $RH^*(A,A)$ be subalgebra in $H^*(A,A)$ generated by $D_i$ and $Z(A)$.
\begin{proposition}\label{P:22}
Inclusion $A\subset \cF$ defines embeddings $H_*(A,A)\to H_*(\cF,\cF)$ and $RH^*(A,A)\to H^*(A,\cF) \cong  H^*(\cF,\cF)$.
\end{proposition}
\begin{proof}
By Proposition 1.9  \cite{GG} homology classes from $H_*(A,A)$ are in one-to-one correspondence with linear combinations of $c_R$ (see Remark \ref{R:cycles}) with coefficients in center $Z(A)$. 
Pairing (\ref{E:pairing}) $H^i(A,A)\cap H_i(A,A)\to H_{0}(A,A)=Z(A)$ between  such such cycles with monomial coefficients from $Z(A)$ and cocycles $D_{i_1}\cup \dots \cup D_{i_s}$ also with monomial coefficients from $Z(A)$ is not degenerate. This can be seen by straightforward application of formula \ref{E:cap}. The pairing  (\ref{E:pairing}) is compatible with embedding $A\subset \cF$. As $A/[A,A]\to \cF/[\cF,\cF]$ is an embedding we conclude that from nondegeneracy of the pairing the maps $H_i(A,A)\to H_i(\cF,\cF)$ and $RH^i(A,A)\to H^i(\cF,\cF)$ are an embedding.
%By Proposition \ref{pc} Hochschild 
\end{proof}

It follows from Proposition \ref{P:restrictioniso} that any derivation of $A$ can be extended in a unique way to derivation of $\cF$.We  think about $D_{x_i}$ as elements of $H^1(\cF,\cF)$.
%\end{group}
\end{document}